\documentclass[aps,showpacs]{revtex4}
\begin{document}
\title{Purely electromagnetic spacetimes}
\author{B.V.Ivanov}
\email{boyko@inrne.bas.bg}
\affiliation{Institute for Nuclear Research and Nuclear Energy,\\
Bulgarian Academy of Sciences\\
Tzarigradsko Shausse 72, Sofia 1784, Bulgaria}

\begin{abstract}
Electrovacuum solutions devoid of usual mass sources are classified in the
case of one, two and three commuting Killing vectors. Three branches of
solutions exist. Electromagnetically induced mass terms appear in some of
them.
\end{abstract}

\maketitle

\section{Introduction}

In general relativity electromagnetic fields alter the metric of spacetime
through their energy-momentum tensor 
\begin{equation}
T_\nu ^\mu =-\frac 1{4\pi }\left( F^{\mu \alpha }F_{\nu \alpha }-\frac 14%
\delta _\nu ^\mu F^{\alpha \beta }F_{\alpha \beta }\right)  \label{one}
\end{equation}
where 
\begin{equation}
F_{\mu \nu }=\partial _\mu A_\nu -\partial _\nu A_\mu  \label{two}
\end{equation}
is the electromagnetic tensor and $A_\mu $ is the four-potential. $T_\nu
^\mu $ enters the r.h.s. of the Einstein equations 
\begin{equation}
R_\nu ^\mu =-8\pi T_\nu ^\mu .  \label{three}
\end{equation}
We use relativistic units $G=$ $c=1$ and take into account that $T_{\mu \nu
} $ is traceless. In addition, the Maxwell equations are coupled to gravity
through the covariant derivatives of $F_{\mu \nu }$%
\begin{equation}
F_{\quad ;\nu }^{\mu \nu }=\frac 1{\sqrt{-g}}\left( \sqrt{-g}F^{\mu \nu
}\right) _\nu =0.  \label{four}
\end{equation}
Here $g$ denotes the metric's determinant and we discuss electrovacuum
solutions. The Einstein-Maxwell (EM) equations (3,4) show how the
electromagnetic field leaves its imprint on the metric.

Already in 1925 Rainich gave the necessary and sufficient conditions for a
gravitational field to originate from a non-null electromagnetic field \cite
{one},\cite{two},\cite{three}. They split into an algebraic and an analytic
part, which involve the Ricci tensor and determine the associated
electromagnetic field up to a constant duality rotation. This method is
cumbersome for finding exact solutions, but is useful to check those already
found \cite{three}. In fact, it is very hard to find any exact solution for
a completely general metric. One has to introduce some symmetry, study
algebraically special fields or fields containing special vectors and
tensors. This seems to simplify the Rainich conditions too. Thus, in Ref. 
\cite{four} it is shown that type D aligned EM solutions are characterized
just by algebraic restrictions on $R_{\mu \nu }$.

In this paper we study the effect on spacetime of electromagnetic fields
possessing some symmetry. The metric inherits this symmetry and has Killing
vectors which commute and increase from one to three. Hence, it depends on
three, two and one coordinates.

When the electromagnetic field is turned off Eq (4) becomes trivial, while
Eq (3) transforms into the vacuum Einstein equations. They have a number of
non-trivial solutions besides flat Minkowski spacetime. Usually
singularities are present where hidden mass sources lie. Thus the
Schwarzschild solution, although being a vacuum one, represents the field of
a point mass, sitting at the origin. We want to study the pure
electromagnetic effect on the metric, therefore we demand that no
traditional masses are present and when the electromagnetic field is turned
off, flat spacetime results. We call such spacetimes purely electromagnetic
spacetimes (PES).

In Section II stationary solutions are discussed by introducing the Ernst
potential, the EM equations based on it and their group of symmetry
transformations. It is shown that PES are characterized by a real constant
taking three values. The metric components are expressed through the
electromagnetic potential, which satisfies one fundamental equation with
three branches of solutions. Sometimes linear terms in the main metric
function appear, leading to electromagnetically induced mass terms.

In Section III static PES with one Killing vector are further elaborated.
There is a large class of solutions based on a harmonic function.

In Section IV stationary axisymmetric PES are discussed with emphasis on
their multipole structure.

In Section V the static subcase is studied and the emergence of Weyl
solutions is clarified.

Section VI is dedicated to PES with plane and cylindrical symmetry. All such
spacetimes are found explicitly, based on a simple harmonic function.

Section VII contains conclusions and discussion.

\section{Stationary PES}

Let the metric possess one Killing vector, taken time-like for convenience.
Such gravitational fields are called stationary \cite{three},\cite{five} and
the interval reads 
\begin{equation}
ds^2=f\left( dx^0+\omega _mdx^m\right) ^2-f^{-1}\gamma _{mn}dx^mdx^n.
\label{five}
\end{equation}
The metric components are independent of time. There exists an Ernst
potential $E$ \cite{six} 
\begin{equation}
E=f-\Psi \Psi ^{*}+i\psi ,\qquad \Psi =\phi +i\lambda ,  \label{six}
\end{equation}
where the scalar electric and magnetic potentials are defined as 
\begin{equation}
F_{0n}=\phi _{,n},\qquad F^{mn}=f\gamma ^{-1/2}\varepsilon ^{mnp}\lambda
_{,p}.  \label{seven}
\end{equation}
The three-dimensional metric $\gamma _{mn}$ determines a corresponding
vector calculus and $\gamma $ is its determinant, while the comma denotes a
derivative. The imaginary part of $E$ follows from $\omega _m,\Psi $ and $f$%
\begin{equation}
\nabla \psi =\vec \tau +i\left( \Psi ^{*}\nabla \Psi -\Psi \nabla \Psi
^{*}\right) ,  \label{eight}
\end{equation}
\begin{equation}
f^{-2}\tau ^m=-\gamma ^{-1/2}\varepsilon ^{mpq}\partial _p\omega _q.
\label{nine}
\end{equation}
The existence of the Ernst potential leads to considerable simplification of
the EM equations for $E,\Psi $ and $\gamma _{mn}$%
\begin{equation}
f\nabla ^2E=\nabla E\left( \nabla E+2\Psi ^{*}\nabla \Psi \right) ,
\label{ten}
\end{equation}
\begin{equation}
f\nabla ^2\Psi =\nabla \Psi \left( \nabla E+2\Psi ^{*}\nabla \Psi \right) ,
\label{eleven}
\end{equation}
\begin{equation}
-f^2R_{mn}\left( \gamma \right) =\frac 12E_{,(m}E_{,n)}^{*}+\Psi E_{,(m}\Psi
_{,n)}^{*}+\Psi ^{*}E_{,(m}^{*}\Psi _{,n)}-\left( E+E^{*}\right) \Psi
_{,(m}\Psi _{,n)}^{*}.  \label{twelve}
\end{equation}
Symmetrization is meant on the r.h.s. of the last equation, while $f$ is
given by Eq (6) 
\begin{equation}
f=\frac 12\left( E+E^{*}\right) +\Psi \Psi ^{*}.  \label{thirteen}
\end{equation}

It is well-known that the group of symmetry transformations of these
equations is $SU\left( 2,1\right) $ \cite{seven}. A non-linear
representation of it consists of 5 transformations containing 3 complex and
2 real parameters \cite{three}, p 520. We shall use two of them that do not
change the solution in a non-trivial manner, namely 
\begin{equation}
E^{\prime }=E+ib,\qquad \Psi ^{\prime }=\Psi ,  \label{fourteen}
\end{equation}
\begin{equation}
E^{\prime }=\alpha \alpha ^{*}E,\qquad \Psi ^{\prime }=\alpha \Psi ,
\label{fifteen}
\end{equation}
while $\gamma _{mn}$ is not transformed. Here $\alpha $ is a complex
parameter, while $b$ is real. The first transformation is a gauge one. The
second is a duality rotation when $\mid \alpha \mid =1$ and a rescaling of $%
ds$ otherwise.

A linear representation of the group of motions is given as follows \cite
{seven}. The Ernst and the electromagnetic potentials are parameterized by 3
complex scalar fields $u,q,w$%
\begin{equation}
E=\frac{u-w}{u+w},\qquad \Psi =\frac q{u+w}.  \label{sixteen}
\end{equation}
One of them ($w$) is redundant and is chosen so that Eqs (10,11) become 
\begin{equation}
\left( uu^{*}+qq^{*}-ww^{*}\right) \nabla ^2Z=2\left( u^{*}\nabla
u+q^{*}\nabla q-w^{*}\nabla w\right) \nabla Z,  \label{seventeen}
\end{equation}
where $Z=u,q$ or $w$. One can further set $u=1,w=\xi $ or $u=\xi ,w=1$
obtaining 
\begin{equation}
E=\frac{1-\xi }{1+\xi }\text{ \qquad or}\qquad E=\frac{\xi -1}{\xi +1}%
,\qquad \Psi =\frac q{1+\xi }.  \label{eighteen}
\end{equation}

The potential $\Psi $ is defined up to a constant $c$. We demand that when $%
\Psi \rightarrow c$ Minkowski spacetime should result. This means that $%
\omega _m=0$ and Eqs (8,9) yield $\psi =0$. Obviously $f$ should become
equal to unity. Then Eq (6) gives $E=1-cc^{*}$, which is a real constant,
containing no mass parameters. When $\Psi $ is turned on, $E$ remains
constant for PES. This is because the real and imaginary part of $\xi $ give
the mass and rotation potentials in the multipole structure of the solution 
\cite{eight}, hence, for PES $\xi $ and correspondingly $E$ should be
trivial constants. We can make them real by transformation (14). Using Eq
(15) the real constant $E_0$ can be set to one of the 3 distinct values $%
1,0,-1$. Hence, there are three branches of PES. In a similar way, the
inequivalent classes of solutions under the action of the complete $SU\left(
2,1\right) $ group are given by $E=1,0,-1$ and $E=E^{*}$ \cite{three}, Fig.
34.1.

The setting of $E=E_0,\psi =0$ for PES causes drastic simplification in the
EM equations. Eq (10) becomes trivial, Eq (13) gives 
\begin{equation}
f=E_0+\Psi \Psi ^{*}.  \label{nineteen}
\end{equation}
The quadratic dependence of $f$ on $\Psi $ can be traced to the quadratic
dependence of the energy-momentum tensor on $F_{\mu \nu }$. Inserting Eq
(19) into Eq (11) one gets an equation for $\Psi $%
\begin{equation}
\left( \Psi \Psi ^{*}+E_0\right) \nabla ^2\Psi =2\Psi ^{*}\nabla \Psi \nabla
\Psi .  \label{twenty}
\end{equation}
Next, $\omega _q$ is determined from Eqs (8,9) 
\begin{equation}
f^2\gamma ^{-1/2}\varepsilon ^{mpq}\partial _p\omega _q=i\left( \Psi
^{*}\nabla \Psi -\Psi \nabla \Psi ^{*}\right) =i\Psi \Psi ^{*}\nabla \ln 
\frac \Psi {\Psi ^{*}}.  \label{twentyone}
\end{equation}
Eq (12), which determines the three-metric $\gamma _{mn}$ simplifies
considerably 
\begin{equation}
f^2R_{mn}\left( \gamma \right) =2E_0\left( \Psi _{,m}\Psi _{,n}^{*}+\Psi
_{,n}\Psi _{,m}^{*}\right) .  \label{twentytwo}
\end{equation}

The 3 branches of the solution are marked by different values of $c=c_1+ic_2$%
\begin{equation}
c_1^2+c_2^2=1-E_0.  \label{twentythree}
\end{equation}
Let us define the potential $\Psi _1=\phi _1+i\lambda _1=\Psi -c$ so that $%
\Psi _1\rightarrow 0$ always when the electromagnetic field is turned off.
Eq (22) remains the same but with $\Psi $ replaced by $\Psi _1$. Eq (19)
becomes 
\begin{equation}
f=1+2c_1\phi _1+2c_2\lambda _1+\phi _1^2+\lambda _1^2.  \label{twentyfour}
\end{equation}
There are obviously linear terms in $f$ whenever $c\neq 0,$ i.e. $E_0\neq 1$.

Neither of the parameterizations (18) can encompass all three branches of
the solution, but this is possible in the parameterization (16) 
\begin{equation}
E_0=-1;\qquad u=0,w=1,\Psi =q,  \label{twentyfive}
\end{equation}
\begin{equation}
E_0=0;\qquad u=w=1,\Psi =q/2,  \label{twentysix}
\end{equation}
\begin{equation}
E_0=1;\qquad u=1,w=0,\Psi =q.  \label{twentyseven}
\end{equation}
In all three cases Eq (17) becomes identical to Eq (20). When $E_0=-1$ Eq
(20) coincides in form with the vacuum Ernst equation for $\xi $ replaced by 
$q$. When $E_0=1$ this equation has been derived also by Tanabe \cite{nine}
from the condition for a linear relation between $u,q,w$. When $E_0=0$ Eq
(20) is equivalent to 
\begin{equation}
\Psi =H^{-1},\qquad \nabla ^2H=0.  \label{twentyeight}
\end{equation}
We also get $R_{mn}\left( \gamma \right) =0$, so that space turns flat, $H$
decouples from $\gamma _{mn}$ and becomes an arbitrary complex harmonic
function. Furthermore 
\begin{equation}
f=\frac 1{HH^{*}},\qquad \varepsilon ^{mpq}\partial _p\omega
_q=iHH^{*}\nabla \ln \frac{H^{*}}H.  \label{twentynine}
\end{equation}
Let $H=L+iM,$ where $L$ and $M$ are real harmonic functions. Then 
\begin{equation}
f=\frac 1{L^2+M^2},\qquad \Psi =\frac 1{L+iM},  \label{thirty}
\end{equation}
\begin{equation}
\phi =\frac L{L^2+M^2}=fL,\qquad \lambda =-\frac M{L^2+M^2}=-fM.
\label{thirtyone}
\end{equation}

Electromagnetic fields alone are unable to induce strong gravitational
fields like those around black holes, therefore $f\approx 1$ and $\phi
,\lambda $ are almost harmonic, as follows from Eq (31). At infinity a
monopole term in $\phi ,\phi \sim e/R$ where $e$ is the charge and $%
R^2=r^2+z^2,$ will induce a long-range mass-type term in $f$ according to Eq
(24) 
\begin{equation}
f\sim 1+2c_1\frac eR  \label{thirtytwo}
\end{equation}
as long as $c_1\neq 0$ ($E_0\neq 1$). It presents an electromagnetically
induced mass. Otherwise the gravitational field of PES remains short-ranged.

Eq (23) does not determine the signs of $c_1$ and $c_2$. One can choose them
in such a way that the induced mass is always positive, no matter what the
sign of the charge $e$ is.

The formulas for the $E_0=0$ branch resemble those of Perj\'es-Israel-Wilson
(PIW) fields \cite{five},\cite{ten},\cite{eleven} but their physical
interpretation is completely different. PIW solutions are derived upon the
condition of flat space metric, which leads to a linear relation between $E$
and $\Psi $ \cite{five} 
\begin{equation}
\Psi =\frac 12\left( 1-E\right) .  \label{thirtythree}
\end{equation}
A comparison with Eq (18) gives $\xi =q$. Thus $\xi $ is not trivial and
there are rotating masses in the system, besides the electromagnetic field.
In fact, the masses of the sources are equal to the charges $m_i=e_i$.
Obviously such solutions are not PES. Curiously, PIW metrics also become
flat when $\Psi $ is turned off, but this is ensured by taking away some
mass in order to keep the relations $m_i=e_i$ intact, even when $%
e_i\rightarrow 0$.

\section{Static PES with one Killing vector}

Stationary gravitational fields are static when the Killing vector is
hyper-surface orthogonal, $\omega _m=0$. Let us write 
\begin{equation}
\Psi =\phi +i\lambda =\chi \cos \delta +i\sin \delta .  \label{thirtyfour}
\end{equation}
Then due to Eq (21) $\delta $ is constant and $\phi \sim \lambda $. The
effects of electric and magnetic fields on gravity are identical and we put
for simplicity $\lambda =0$. The same conclusion holds for general static
fields \cite{twelve},\cite{thirteen},\cite{fourteen}. This choice makes $%
\Psi $ real and equal to $\phi $. One can always maintain that it depends on
the polar coordinates $r,z,\varphi $ through some other function $h$. Eq
(20) becomes 
\begin{equation}
\left[ \left( \phi ^2+E_0\right) \phi _{hh}-2\phi \phi _h^2\right] \nabla
h\nabla h=-\left( \phi ^2+E_0\right) \phi _h\nabla ^2h.  \label{thirtyfive}
\end{equation}
A large class of solutions may be found when $h$ is harmonic. Then Eq (35)
can be integrated 
\begin{equation}
h=\int \frac{d\phi }{\phi ^2+E_0}.  \label{thirtysix}
\end{equation}
The integral has 3 analytic expressions according to the value of $E_0$ and
in all of them the functional dependence $h\left( \phi \right) $ can be
inverted 
\begin{equation}
E_0=0,\qquad \phi =-1/h,  \label{thirtyseven}
\end{equation}
\begin{equation}
E_0=1,\qquad \phi =\tan h,  \label{thirtyeight}
\end{equation}
\begin{equation}
E_0=-1,\qquad \phi =\frac{1+e^{2h}}{1-e^{2h}}.  \label{thirtynine}
\end{equation}
In the last case we took into account that $c=\sqrt{2}>1$. The denominator
in Eq (36) is in fact $f,$which leads to the relation $\phi _{,i}=fh_{,i}$
similar to Eq (31). Thus $\phi $ is almost harmonic. Eq (22) becomes 
\begin{equation}
R_{mn}\left( \gamma \right) =\frac{4E_0\phi _m\phi _n}{\left( \phi
^2+E_0\right) ^2}.  \label{forty}
\end{equation}
In the branch $E_0=0$ space is flat, $h$ decouples from $\gamma _{mn}$ and
the solution resembles the Majumdar-Papapetrou (MP) solutions \cite{fifteen},%
\cite{sixteen}, but still the physical interpretation is different. The
latter solutions are a static subcase of the PIW solutions and their sources
have masses equal to the charges.

\section{Stationary axisymmetric PES}

Such metrics possess two commuting Killing vectors and do not depend on $t$
and $\varphi $. The interval is 
\begin{equation}
ds^2=f\left( dt-\omega d\varphi \right) ^2-f^{-1}\left[ e^{2k}\left(
dr^2+dz^2\right) +r^2d\varphi ^2\right] .  \label{fortyone}
\end{equation}

The solutions with $E_0=\pm 1$ in the previous sections are inexplicit
because $\nabla $ and $\nabla ^2$ depend on $\gamma _{mn},$ which depends in
turn on $\Psi $ through Eqs (12,22,40). This vicious circle breaks for all
axisymmetric solutions; the gradient and the Laplacian are the flat
3-dimensional ones. Eq (20) decouples from $k$. The same is true for Eq (21) 
\begin{equation}
\frac{f^2}r\omega _r=2\left( \phi \lambda _z-\lambda \phi _z\right) ,\qquad 
\frac{f^2}r\omega _z=2\left( \lambda \phi _r-\phi \lambda _r\right) .
\label{fortytwo}
\end{equation}
Eq (19) is unchanged, while Eq (22) simplifies to 
\begin{equation}
k_r=\frac 12r\left( R_{rr}-R_{zz}\right) =\frac{2E_0r}{f^2}\left( \Psi
_r\Psi _r^{*}-\Psi _z\Psi _z^{*}\right) ,  \label{fortythree}
\end{equation}
\begin{equation}
k_z=rR_{rz}=\frac{2E_0r}{f^2}\left( \Psi _r\Psi _z^{*}+\Psi _z\Psi
_r^{*}\right) .  \label{fortyfour}
\end{equation}
The equations for $\omega $ and $k$ are linear and can be easily solved
after $\Psi $ is found from Eq (20).

When $E_0=0$ Eqs (28-31) still hold and together with Eq (42) give the PES
solution. As an example, let us choose $L$ and $M$ as 
\begin{equation}
L=1+\frac eR,\qquad M=\frac{\mu z}{R^3},  \label{fortyfive}
\end{equation}
\begin{equation}
\Psi =\frac{R^3}{R^3+eR^2+i\mu z}.  \label{fortysix}
\end{equation}
Then near infinity we have 
\begin{equation}
\phi \sim 1-\frac eR,\qquad \lambda \sim -\frac{\mu z}{R^3},\qquad f\sim 1-%
\frac{2|e|}R,  \label{fortyseven}
\end{equation}
so that the solution has charge $-e$, magnetic moment $-\mu $ and a monopole
term of purely electromagnetic origin in $f$, corresponding to a mass equal
to $|e|$. The angular momentum is $-\mu $ and is of magnetic origin. This
solution represents a massive charged magnetic dipole, whose mass and
angular momentum are induced electromagnetically.

There are different techniques for solving the EM equations in the general
axisymmetric case. One of them is the Sibgatullin-Manko method, based on a
prescribed behavior of the Ernst potential on the symmetry axis \cite
{seventeen},\cite{eighteen}. One starts with 
\begin{equation}
E\left( r=0,z\right) =1+\sum_{l=1}^N\frac{\alpha _l}{z-\beta _l}
\label{fortyeight}
\end{equation}
and a similar expression for $\Psi $, where $\alpha _l,\beta _l$ are given
constant parameters related to the mass, angular momentum and higher
multipole moments. The solution is obtained by a sophisticated integration
procedure. It is characterized by the gravitational and electromagnetic
moments $P_i$ and $Q_i$ \cite{eight},\cite{nineteen}. They are determined
from the coefficients $\xi _i,q_i$ of the series expansion of $\xi $ and $q$
near infinity. The infinite point is brought to the origin of the
coordinates by a conformal transformation $\left( r,z\right) \rightarrow
\left( \bar r,\bar z\right) $. One can write 
\begin{equation}
\xi \left( \bar r=0,\bar z\right) =\sum_{i=0}^\infty \xi _i\bar z%
^{i+1},\qquad q\left( \bar r=0,\bar z\right) =\sum_{i=0}^\infty q_i\bar z%
^{i+1},  \label{fortynine}
\end{equation}
where $\bar z=1/z$. Eq (48) shows that PES is obtained as a special massless
case when $\alpha _l=\beta _l=0$ and this gives always a member of the $%
E_0=1 $ branch. The same conclusion follows from Eq (49); $\xi $ becomes
trivial only when all $\xi _i=0$. Then $\xi =0$ and $E=1$. The massless case
has been discussed in a number of papers \cite{twenty},\cite{twentyone},\cite
{twentytwo},\cite{twentythree}. In the last reference the explicit solution
for a massless magnetic dipole is presented. One can check that in all these
examples $E=1$ and therefore $c=0$. There are no linear terms in $f$ and no
electromagnetically induced mass.

It must be pointed out that the expansion in Eq (49) cannot include $\xi =1$
and consequently the $E_0=0$ branch of PES. This happens because there is no
constant term in it. On the other side PIW solutions have $\xi _i=q_i$ and
do fit into Eq (49). In addition, they are massive, not massless solutions.

The branch $E_0=-1$ has $\Psi =q$ and Eq (20) is equivalent to the vacuum
Ernst equation for $q$ instead of $\xi $. Its solutions can be found by the
Sibgatullin-Manko method and many other methods \cite{three}. The metric
components $f,\omega $ and $k$ are given by Eqs (19,42-44).

\section{Static axisymmetric PES}

Based on the previous sections one can describe these spacetimes as follows.
The function $\omega $ vanishes and $\phi \sim \lambda $. We put $\lambda =0$
for simplicity. Eq (20) becomes 
\begin{equation}
\left( \phi ^2+E_0\right) \nabla ^2\phi =2\phi \nabla \phi \nabla \phi
\label{fifty}
\end{equation}
where the differential operators are flat and 3-dimensional. Next 
\begin{equation}
f=1+2c_1\phi _1+\phi _1^2=E_0+\phi ^2  \label{fiftyone}
\end{equation}
where $E_0=1-c_1^2,\phi =\phi _1+c_1$ and $\phi _1\rightarrow 0$ when the
electric field is turned off and also at infinity. Finally 
\begin{equation}
k_r=\frac{2E_0r}{f^2}\left( \phi _r^2-\phi _z^2\right) ,\qquad k_z=\frac{%
4E_0r}{f^2}\phi _r\phi _z.  \label{fiftytwo}
\end{equation}
The class of solutions given by Eq (36) is still valid, $h$ being an
arbitrary real harmonic function. It was discovered by H.Weyl in 1917 \cite
{twentyfour} and includes the three branches (37-39). The branch $E_0=0$ is
exhausted by Weyl solutions and is conformastatic, but for $E_0=\pm 1$ there
can be solutions of Eq (50) not based on a harmonic function. Similarly to
the stationary axisymmetric PES with $E_0=1$, static axisymmetric PES with
this property are among the massless cases of static electrovacuum
Sibgatullin-Manko solutions \cite{twentyfive}. There are of course
electrovacuum solutions containing usual mass sources, but the quadratic
relation in Eq (51) breaks for them in principle.

Weyl solutions are usually derived by demanding functional dependence $%
f\left( \phi _1\right) $ which leads to $f_{\phi _1\phi _1}=2$ and leaves
the coefficient $c_1$ in Eq (51) undetermined. We see that it descends from
the Ernst potential $E_0$ of the solution and is fixed to one of three
distinct values $0,1,\sqrt{2}$.

\section{PES with 3 commuting Killing vectors}

Such metrics depend on just one coordinate; $z$ (plane symmetry) or $r$
(cylindrical symmetry). More generally, they can depend on some function of $%
r,z$ and spherical symmetry is also included, but we shall not discuss this
possibility in the present paper. Eq (42) tells us that when $\Psi =\Psi
\left( z\right) $ it induces $\omega \left( r\right) $ and vice versa. The
symmetry of the electromagnetic field is inherited only when $\omega =0$ and
the solution is static. There are stationary cylindrical solutions like the
one given by Eq (22.17) from Ref \cite{three} but it doesn't transfer into
Minkowski spacetime when the electric field is turned off and is not a PES.
Now $\phi $ is a function of either $z$ or $r$ and consequently is a
function of either $h=1-Qz$ or $h=1-Q\ln r$ which are harmonic. Thus Eq (20)
transforms into Eq (36) and its 3 branches comprise all solutions. Eq (52)
becomes 
\begin{equation}
k_r=2E_0r\left( h_r^2-h_z^2\right) ,\qquad k_z=4E_0rh_rh_z.
\label{fiftythree}
\end{equation}

Let us discuss first the plane-symmetric case. We have $k_z=0$ and 
\begin{equation}
k=-E_0Q^2r^2.  \label{fiftyfour}
\end{equation}
The plane symmetry of the electric field is not inherited by the metric
unless the branch $E_0=0$ is chosen. Then $k=0$ and 
\begin{equation}
\phi =\frac 1{1-Qz},\qquad f=\left( 1-Qz\right) ^{-2}.  \label{fiftyfive}
\end{equation}
This is written in the form of Eq (51) like 
\begin{equation}
f=1+2\phi _1+\phi _1^2,\qquad \phi _1=\phi -1,  \label{fiftysix}
\end{equation}
so that when $Q=0$ we get $\phi _1=0$ and $f=1$. There is clearly a linear
term in $f$.

The electrified plane-symmetric solution has been studied by many authors,
working in different coordinate systems. It appeared for the first time in a
paper by Kar \cite{twentysix}. The case with $k=0$ was singled out by
McVittie \cite{twentyseven}.

In the cylindrical case one should replace $z$ by $\ln r$ in Eq (55). The
function $k=k\left( r\right) $ again, but now all three branches preserve
cylindrical symmetry. The one with $E_0=0$ is the metric given by Eq (22.16)
from \cite{three} after some corrections are made. It was found by Bonnor 
\cite{twentyeight} and rederived by Raychaudhuri \cite{twentynine} who used
the Rainich formalism.

\section{Conclusions and discussion}

We have shown in this paper that the Rainich program of describing
electromagnetically induced metrics can be implemented in a simpler way and
in more detail when the symmetry of the system is gradually increased. PES
require that the Ernst potential becomes a constant with 3 possible values.
This invokes a quadratic dependence of the main metric function $f$ on the
electromagnetic potential $\Psi $ and its complex conjugate. The two basic
field equations (10,11) reduce to a single one Eq (20) for $\Psi $. It
possesses 3 branches of solutions - one harmonic, one of Ernst type and one
of quasi-Ernst type. In some cases $f$ has linear in $\Psi _1$ terms, which
give rise to electromagnetically induced mass terms. Solutions in harmonic
functions form a large class when one or two Killing vectors are present and
become exhaustive when the metric depends on a single coordinate. We have
also shown that the 3-branched Weyl solution is a kind of PES and has
analogs in axisymmetric and stationary fields.

There are many EM solutions in the literature and it seems strange that PES
were not studied systematically in the past. The reason probably is that
vacuum solutions are studied first as being simpler. Then they are
electrified and magnetised. In this way traditional mass sources co-exist
with the electromagnetic field and the resulting gravitation is a mixture
due to the both types of sources. In other areas of general relativity the
corresponding problem has already been investigated.

In the case of charged perfect fluids there is a bunch of spherically
symmetric models where mass arises in a purely electromagnetic way \cite
{thirty},\cite{thirtyone}. Some of them were proposed as classical models of
the electron.

The study of the gravitation of beams of incoherent light, whose
energy-momentum tensor is of pure radiation type, began already in 1931 in
the linear approximation \cite{thirtytwo}. Later Bonnor found exact
solutions belonging to the class of pp-waves \cite{thirtythree}. Even the
gravitational field of two identical colliding beams of light was found
recently \cite{thirtyfour}. Although these fields are rather weak, such
studies have important conceptual motive to unravel the non-linear structure
of general relativity.

The sources of PES are point mixtures of electric and magnetic multipoles
without any mass multipoles. There should also be regular PES arising from
rotating charged and magnetised surfaces. In this respect they are
completely different from PIW and MP solutions whose sources are massive and
represent extreme black holes or shells of charged dust balanced by masses
equal to the charges \cite{thirtyfive},\cite{thirtysix},\cite{thirtyseven}.

\end{document}